\documentclass[twocolumn,aps,tightenlines,floatfix,showpacs]{revtex4}
\usepackage{bm}
\usepackage{graphics}
\usepackage{graphicx}
\usepackage{epsfig}
\usepackage{longtable}
\newcommand{\btab}{\begin{tabular}}
\newcommand{\etab}{\end{tabular}}
\usepackage{multirow,amsmath,booktabs}
\begin{document}

\title{Tensor interaction contributions to single-particle energies}

\author{B. A. Brown$^{1}$, T. Duguet$^{1}$, T. Otsuka$^{1,2}$, D. 
Abe$^{3}$
and T. Suzuki$^{4}$}

\affiliation{1) Department of Physics and Astronomy
and National Superconducting
Cyclotron Laboratory,
Michigan State University,
East Lansing, Michigan 48824-1321, USA}
\affiliation{2) Department of Physics and Center for Nuclear Study,
University of
Tokyo, Hongo, Tokyo 113-0033, Japan; and RIKEN,
Hirosawa, Wako-shi, Saitama 351-0198, Japan}
\affiliation{3) Department of Physics,
University of
Tokyo, Hongo, Tokyo 113-0033, Japan}
\affiliation{4) Department of Physics,
Nihon University, Sajurajosui, Setagay-ku, Tokyo 156-8550, Japan}

\begin{abstract}
We calculate the contribution of the nucleon-nucleon tensor
interaction to single-particle energies with
finite-range $  G  $ matrix potentials and with zero-range Skyrme
potentials. The Skx Skyrme parameters including
the zero-range tensor terms with strengths calibrated to the
finite-range results are refitted to nuclear properties.
The fit allows the zero-range proton-neutron tensor interaction
as calibrated to the finite-range potential results and
that gives the observed change in the single-particle
gap $\epsilon$(h$_{11/2}$)-$\epsilon$(g$_{7/2}$)
going from $^{114}$Sn to $^{132}$Sn. However, the
experimental $\ell$ dependence
of the spin-orbit splittings in $^{132}$Sn and $^{208}$Pb
is not well described when the tensor is added, due to
a change in the radial dependence of the total spin-orbit
potential. The gap shift and a good fit to the
$\ell$-dependence can be recovered when the like-particle
tensor interaction is opposite in sign to that required
for the $  G  $ matrix.

\end{abstract}
\date{\today}
\maketitle

The tensor force between nucleons is important for
the single-particle energies spacing and the
shell structure of nuclei obtained from shell-model
configuration interaction models of nuclei \cite{ot1},
\cite{ot2}.
But, except for an early exploratory work \cite{stancu},
its role in Hartree-Fock models has been neglected.
Results for relative energy shifts have recently
been obtained with a finite-range tensor in the
Gogny model \cite{gogt}. In this letter we present the
first systematic results for a Skyrme-type
interaction with a zero-range tensor interaction
from a global fit to nuclear data including data
for single-particle energies. We start by
calibrating the strength of the zero-range tensor with
results obtained from a $  G  $ matrix interaction.
Then the Skyrme parameters plus the
tensor parameters are varied to
obtain a best fit to data. We find that the
fit will allow an isovector tensor strength
similar to that expected from the $  G  $ matrix.
But the data set prefers an isoscalar tensor
that is much smaller than expected from the
$  G  $ matrix. The reason is traced to a difference
in between the Skyrme spin-orbit and tensor
radial forms in the doubly closed-shell nuclei
$^{132}$Sn and $^{208}$Pb.

First we calculated the contribution to the single-particle
proton energies for single-particle states
above the $  Z=50  $ closed shell from the tensor part of the
Hosaka-Kubo-Toki (HKT) $  G  $ matrix \cite{hkt}. HKT is a
one-boson exchange potential that reproduces
the $  G  $ matrix elements obtained
from the Paris potential. This tensor interaction
has the form:
$$
 V^{t} = S_{12}
  \displaystyle\sum _{i,T} W_{i,T}
   \biggl\{ 1+\frac{3}{x_{i}}+\frac{3}{x_{i}^{2}} \biggr\}
   \frac{e^{-x_{i}}}{x_{i}}       \eqno({1})
$$
where
$$
 S_{12} = 3(\vec{\sigma}_{1} \cdot \hat{r})
   (\vec{\sigma}_{2} \cdot \hat{r}) - (\vec{\sigma}_{1} \cdot 
\vec{\sigma}_{2})
= Y^{(2)}(\hat{r}) \cdot \sqrt{ \frac{24\pi }{5}}\, [\vec{\sigma}_{1} 
\otimes \vec{\sigma}_{2}]^{(2)},
$$
$  x_{i} = r/r_{i}  $ where r$_{i}$ are the range parameters.
This consists of the one-pion
exchange potential with $  r_{\pi }=1.414  $ fm,
$  [W_{\pi ,T=0}/W_{\pi ,T=1}] = -3  $, and $  W_{\pi ,T=1}=3.49  $ 
MeV, plus short-range
potential with $  r_{s}=0.25  $ fm and with the strengths $  W  $ 
determined
from the $  G  $ matrix elements: $  W_{s,T=0}=3105  $ MeV and
$  W_{s,T=1}=-1382  $ MeV.

The contributions to the proton
single-particle states for $^{132}$Sn are shown in Table I.
They were obtained with harmonic oscillator radial wavefunctions
with $\hbar\omega$=7.87 MeV. (The results with the tensor part of the
M3Y potential \cite{m3y} are the same as HKT within about 3\%).
The contribution to the single-particle
energy of the valence proton
in orbital $  k=(n,\ell ,j)  $ from the core protons is obtained from:
$$
(2j+1) E^{t}_{kp} = \displaystyle\sum _{k',J} (2J+1)
V^{t}_{k,k',J,T=1},       \eqno({2})
$$
and from the core neutrons from:
$$
(2j+1) E^{t}_{kn} = \displaystyle\sum _{k',J,T} \frac{1+\delta 
_{k,k'}}{2} (2J+1)
V^{t}_{k,k',J,T},       \eqno({3})
$$
where the two-body matrix elements are\\
$  V^{t}_{k,k',J,T} =\, <k,k',J,T\mid V^{t}\mid k,k',J,T>  $.

In the sum over the core orbitals $  k'  $ the contributions
from the sum of $  j'_{>}=\ell '+1/2  $ and $  j'_{<}=\ell '-1/2  $ 
orbital
pairs cancel when both are filled
as shown in Eq.\ (4) of \cite{ot2}. Thus the $  E_{kq}^{t}  $
are zero for $  LS  $ closed cores. For non-$  LS  $ closed
cores with
a pair of valence orbits with $  j_{>}=\ell +1/2  $ and $  j_{<}=\ell 
-1/2  $
the energy shifts are
$  (2j_{>}+1) E^{t}_{k_{>}q} = - (2j_{<}+1)E^{t}_{k_{<}q}  $, which 
means that for
a given $\ell$ value the tensor interaction with the core
contributes to the
effective spin-orbit splitting (see Eq.\ (4) of \cite{ot2}).
The short-range contribution ($  s  $) to the energy shifts
are given in the middle part of Table I. One observes the
change in sign that is related to the partial cancellation
of the $\pi$-exchange potential by the short-range potential.
The variation of the ratio over several valence
orbitals shown at the bottom of Table I
is a measure of how well
the finite-range tensor can be approximated by a
zero-range form. This is important since the zero-range
approximation leads to an analytic form for the
tensor density functional, and an efficient
implementation in the Skyrme Hartree-Fock method \cite{stancu}.
The ratio varies depending on
which orbits are filled in the core,
up to a factor of two. But in the case
of total energy for protons in $^{132}$Sn the orbit
dependence in the ratio is small.
\begin{table}
\begin{center}
\caption{Contributions of the tensor finite-range $  G  $ matrix
interaction to single-particle proton energies in $^{132}$Sn.
The results are given for
$  E^{t}_{kp}  $: contribution from the $  0g_{9/2}  $ proton orbital,
$  E^{t}_{kn}(100)  $: contribution from the $  0g_{9/2}  $
 neutron orbital,
$  E^{t}_{kn}(114)  $: $  E^{t}_{kn}(100)  $ plus the contribution from 
the
$  0g_{7/2}  $ and $  1d_{5/2}  $ neutron orbitals,
$  E^{t}_{kn}  $: $  E^{t}_{kn}(114)  $ plus the contribution from the
 $  1d_{3/2}  $ and $  0h_{11/2}  $ neutron orbitals.
The short-range tensor contribution is given by the ``$  s  $-only"
results.}
\begin{tabular}{|r|r|r|r|r|r|r|r|}
\hline
type & $  k=(n\ell _{j})  $ & $  E^{t}_{kp}  $
& $  E^{t}_{kn}(100)  $ & $  E^{t}_{kn}(114)  $
  & $  E^{t}_{kn}  $  & $  E^{t}_{kp}  $ \\
 &  &
& &
  &   & $  +E^{t}_{kn}  $ \\
   &   & (MeV) & (MeV) & (MeV)  & (MeV) & (MeV) \\
\hline
\hline
 $  \pi +s  $  & $  0g_{7/2}  $  & -0.458 & -1.009 & -0.135 &  -1.032  
& -1.490 \\
 & $  1d_{5/2}  $  &  0.078 &  0.180 &  0.395 &   0.218  &  0.296 \\
 & $  1d_{3/2}  $  & -0.118 & -0.270 & -0.593 &  -0.328  & -0.446 \\
 & $  0h_{11/2}  $ &  0.308 &  0.688 &  0.109 &   0.848  &  1.156 \\
\hline
$  s  $ only & $  0g_{7/2}  $  &  0.251 &  0.408 &  0.072 &  0.465   &  
0.716  \\
 & $  1d_{5/2}  $  & -0.060 & -0.097 & -0.162 & -0.100   & -0.160  \\
 & $  1d_{3/2}  $  &  0.090 &  0.145 &  0.243 &  0.150   &  0.240 \\
 & $  0h_{11/2}  $ & -0.191 & -0.310 & -0.050 & -0.397   & -0.588 \\
 \hline
$  \frac{\pi +s}{s}  $ & $  0g_{7/2}  $  & -1.82  & -2.48 & -1.88
         & -2.22   & -2.08 \\
 & $  1d_{5/2}  $  & -1.31  & -1.86 & -2.44 & -2.18   & -1.85 \\
 & $  1d_{3/2}  $  & -1.31  & -1.86 & -2.44 & -2.18   & -1.88 \\
 & $  0h_{11/2}  $ & -1.62  & -2.22 & -2.18 & -2.14   & -1.97 \\
\hline
\end{tabular}
\end{center}
\end{table}

One observes from Table I that the tensor interaction
results in a change in the $  0g_{7/2}  $-$  0h_{11/2}  $ gap going
from $^{114}$Sn (e.g. where only the $  1d_{5/2}  $ and $  0g_{7/2}  $
orbitals are filled) to $^{132}$Sn of 1.64 MeV. $^{132}$Sn is one
of the best doubly magic nuclei, and
the lowest levels in $^{133}$Sb are thus taken
as single-particle states for adding a proton to
$^{132}$Sn, although the experimental measurement
of spectroscopic strength from one-proton transfer reactions
has not yet been carried out. The single-particle
proton energies for $^{132}$Sn are given in Table II.

Proton transfer experiments have been carried out for
$^{114}$Sn to $^{115}$Sb. But the interpretation of
the experimental results in terms of single-particle
energies is not so simple since the neutron configuration
in $^{114}$Sn is not magic with significant configuration mixing
between the lowest neutron orbits of $  0g_{7/2}  $
and $  1d_{5/2}  $ and the upper orbits of
$  1d_{3/2}  $, $  2s_{1/2}  $ and $  0h_{11/2}  $. To estimate
the effect of splitting of single particle strength
we carry out a large-basis shell-model calculations that
includes up to three neutrons being excited from the
lower to the upper orbits with the renormalized $  G  $ matrix
interaction from \cite{babg}. The spectroscopic strength
obtained is shown in figure 1. One observes the lowest
$  J=j  $ states contains the largest fraction of single-particle
strength, but that there is significant spreading to higher energy.
The isolation of one large part of the spectroscopic
strength into the lowest state is consistent with
experimental observation  \cite{schiffer}.
This spreading is due to coupling with the neutron vibrations
within the $  0g_{7/2},1d,2s,h_{11/2}  $ model space
as well as isospin splitting of the strength to
the $  T_{>}=15/2  $ states. The centroid
energies obtained from the results of Fig.\ 1 are within
about 10 keV of the single-particle energies obtained
with the simplest $  [\nu ,0g_{7/2}^{8},1d_{5/2}^{6}][\pi _{n,\ell ,j}] 
 $ configuration.
This simple configuration is the one assumed for the finite-range
tensor contribution discussed above as well as for the Skyrme
Hartree-Fock calculations to be discussed below. (This configuration
does not have good isospin, but configuration mixing restores
isospin.) In order to estimate the proton single-particle
energies in $^{115}$Sb we add on to the separation
energy of the lowest
states of a given $  J=j  $ in the experimental
spectrum a correction based on the
configuration mixing results of Fig.\ 1, giving the experimental
centroid energies in column 3 of Table II.
\begin{figure}
\scalebox{0.5}{\includegraphics{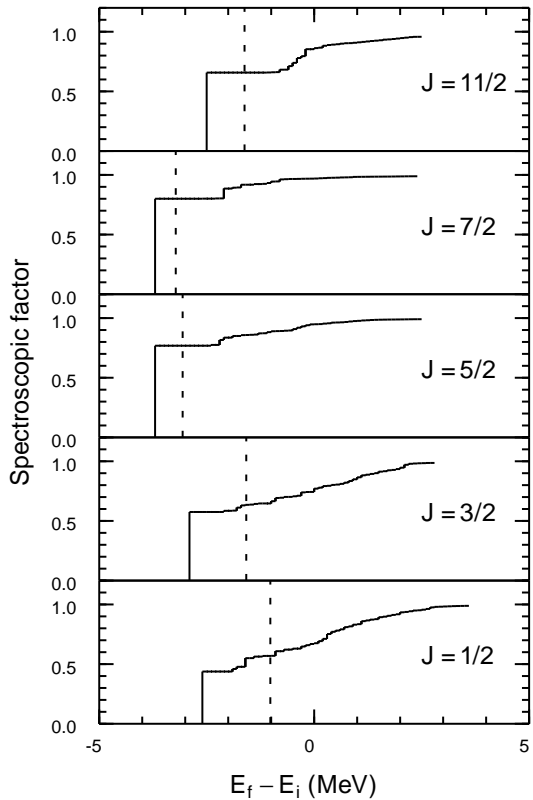}}
\caption{Single-particle proton spectroscopic factors for
$^{114}$Sn to $^{115}$Sb from the shell-model calculations.
The lines for each $  n,\ell ,j  $ value correspond to the
cumulative sum of spectroscopic strength as a
function of the
energy-difference $  E_{f}-E_{i}  $. The centroid energies
are indicated by the dashed lines.}
\label{(1)}
\end{figure}
\begin{table}
\begin{center}
\caption{Proton single-particle energies in $^{114}$Sn and $^{132}$Sn.
The gap is the energy difference between $  0g_{7/2}  $ and $  
0h_{11/2}  $.}
\begin{tabular}{|r|r|r|r|r|r|r|}
\hline
Nucleus & $  n\ell _{j}  $ & exp     & exp      & Skx  & Skxta  & Skxtb 
  \\
       &    & lowest $  J  $   & centroid &      &        & \\
       &    & (MeV) & (MeV) & (MeV)  & (MeV) & (MeV) \\
\hline
\hline
$^{132}$Sn &  $  0g_{7/2}  $   & -9.68 & -9.68 & -9.87  &  -10.82 & 
-9.86 \\
 & $  1d_{5/2}  $           & -8.72 & -8.72 & -9.20  &  -9.22  & -9.30  
\\
 & $  1d_{3/2}  $           & -6.97 & -6.97 & -7.38  &  -7.50  & -7.14  
\\
 & $  2s_{1/2}  $           &       &       & -6.82  &  -6.93  & -6.78  
\\
 & $  0h_{11/2}  $          & -6.89 & -6.89 & -6.92  &  -6.08  & -6.66  
\\
 & gap &  2.79 &  2.79 &  2.84  &   4.52  &  3.20 \\
\hline
$^{114}$Sn &  $  0g_{7/2}  $           & -3.01 & -2.48 & -2.78  &  
-2.70 &  -1.59 \\
 & $  1d_{5/2}  $           & -3.73 & -3.02 & -2.91  &  -2.81 &  -2.66  
\\
 & $  1d_{3/2}  $           & -2.66 & -1.63 & -0.88  &  -1.06 &  -1.16  
\\
 & $  2s_{1/2}  $           & -2.96 & -1.32 & -0.81  &  -0.87 &  -0.85  
\\
 & $  0h_{11/2}  $          & -2.43 & -1.45 &  0.19  &   0.11 &  -0.76  
\\
 & gap  &  0.58 &  1.03 &  2.97  &   2.81 &   0.83 \\
\hline
\end{tabular}
\end{center}
\end{table}

In addition to configuration mixing within the
$  0g_{7/2},1d,2s,h_{11/2}  $ model space, one could
consider the effects of protons excited across the
$  Z=50  $ shell gap and neutrons excited across the
$  N=50  $ and $  N=82  $ shell gaps. The proton excitations
have a direct affect on enhancing the $  B(E2)  $ values
for low-lying 2$^{ + }$ states of Sn \cite{banu}. This type of 
configuration
mixing affects single-particle energies for all of the
Sn isotopes including those for $^{133}$Sb. However,
this is already partly accounted for in Skx
with its empirical effective mass of near unity \cite{skx}.
The enhancement of the
bare $  G  $ matrix effective mass of $    m^{*}/m=0.6-0.7   $
towards its empirical value of near unity in Skx
can be attributed to coupling with the multipole
vibrations of the core \cite{53}, \cite{54}.

The experimental centroid energies are compared with the results
of Skyrme Hartree-Fock calculations. The Skx interaction \cite{skx}
was obtained from a fit to binding energies, rms charge radii
and single-particle energies including those for the $^{132}$Sn
core \cite{skx}. Thus the rather good agreement
between the experimental and Skx proton single-particle
energies for $^{132}$Sn is not an accident. The $^{114}$Sn data
were not included in the original Skx fit
since, as discussed, this does
not have a magic neutron number. However, the agreement with the
centroid energies is not bad except for the $  0h_{11/2}  $.

The gap between the $  0h_{11/2}  $ and $  0g_{7/2}  $ single
particle energies is particularly sensitive to the tensor
interaction. Experimentally the gap changes by 1.76 MeV;
from 2.79 MeV
in $^{132}$Sn to 1.03 MeV in $^{114}$Sn. The gap for Skx
is about the same for $^{114}$Sn and $^{132}$Sn
and this is similar to what we find for other
Skyrme interactions. But the finite-range tensor
interaction discussed above
leads to about a gap shift of 1.64 MeV - close to
the observed shift of 1.76 MeV and to the
values shown in Fig.\ 4(d) of \cite{ot2} and Fig.\ 4
of \cite{gogt}. Thus we are motivated
to add a tensor interaction to the
Skyrme functional.

We use the zero-range form of the tensor potential
given by Stancu et al. \cite{stancu} (Eq.\ 1 of their paper).
The zero-range tensor gives an additional
contribution to the
Skyrme spin-orbit term
of the form:
$$
\Delta W_{n} = \alpha J_{n} + \beta J_{p}, \hspace{1cm} \Delta W_{p} = 
\alpha J_{p} + \beta J_{n}       \eqno({4})
$$
where the coefficients $\alpha$ and $\beta$ (given in units
of MeV fm$^{5}$) come from the
zero-range form of the tensor interaction ($\alpha_{t}$ and $\beta_{t}$)
as well as from the exchange part of the central interaction:
$$
\alpha _{c} = \frac{1}{8}(t_{1}-t_{2}) - \frac{1}{8}(t_{1} x_{1} + 
t_{2} x_{2})       \eqno({5})
$$
and
$$
\beta _{c} = - \frac{1}{8}(t_{1} x_{1} + t_{2} x_{2}).       \eqno({6})
$$
The $  J_{q}  $ are the spin densities defined by
$$
J_{q}(r) = \frac{1}{4\pi r^{3}} \displaystyle\sum _{\alpha } 
(2j_{\alpha }+1)
[j_{\alpha }(j_{\alpha }+1)-\ell _{\alpha }(\ell _{\alpha 
}+1)-\frac{3}{4}] R^{2}_{\alpha }(r)
$$
where the sum is over the occupied orbits with proton
($  q=p  $) or neutrons ($  q=n  $).

We start with the Skx interaction \cite{skx} and the data base that
was used to determine its parameters. Skx is the
only interaction for which a large number of
experimental single-particle energies where used to
constrain the parameters. As a baseline for our new
fits, Skx gives a $\chi^{2}$ value
of 0.60 when the parameters
$  t_{0}, t_{1}, t_{2}, t_{3}, x_{0}, x_{1}, x_{2}, x_{3}  $ and $  W  
$ are
fitted to the data set of \cite{skx}.
For this original Skx
the $\alpha_{c}$ and $\beta_{c}$ terms were not included. If they
are included, the $\chi^{2}$ increases slightly to 0.62
and the central-exchange values are $\alpha_{c}$ = 24 and $\beta_{c}$ = 
-23.

As For the tensor contribution, the
initial set of $\alpha_{t}$ and $\beta_{t}$ parameters were chosen to
reproduce
the calculated $  E^{t}_{pk}  $ and $  E^{t}_{nk}  $
values for the $  0g_{7/2}  $ proton orbit
from the finite-range $  G  $ matrix given
in Table I.
The results are $\alpha_{t}$=60 and $\beta_{t}$=110. These
are larger than Skx central-exchange values
of $\alpha_{c}$ = 24 and $\beta_{c}$ = -23, but both should
be considered for the total and the refits we
carry out will include the effects of $\alpha_{c}$
and $\beta_{c}$. For comparison with Stancu et al.
\cite{stancu},
our values of $\alpha_{t}$ and $\beta_{t}$ are close
to the those they estimate from
the interaction of Sprung and Banerjee \cite{sb}
with $  q=1.0  $ fm$^{-1}$ (Table 1 of \cite{stancu}).

Then the Skyrme parameters
$  t_{0}, t_{1}, t_{2}, t_{3}, x_{0}, x_{1}, x_{2}, x_{3}  $ and $  W  
$ were
refit for these fixed values of $\alpha_{t}$ and $\beta_{t}$
with a resulting set of parameters called Skxta.
The $\chi^{2}$ value
increased significantly to 1.50. The contributions from the
zero-range tensor, central-exchange and spin-orbit interactions to the
proton single-particle energies in $^{132}$Sn are shown in
Table III. The central exchange values from the fit
are $\alpha_{c}$=33 and $\beta_{c}$=-16.

The single-particle
energies for orbitals around $^{132}$Sn obtained with Skx
and Skxta are shown in Figs.\ 2 and 3 respectively. Comparison
of these figures shows that the $\chi^{2}$ increase is due to a
poorer $\ell$-dependence of the spin-orbit splitting for Skxta compared
to Skx. This can be traced to a difference in the radial functional
form of the spin-orbit contributions that are shown in Fig.\ 4.
The tensor contribution peaks at a  0.5 fm smaller radius compared
to the Skyrme spin-orbit potential. Since the tensor
contribution with $\alpha_{t}$=60 and $\beta_{t}$=110
is opposite in sign to the normal spin-orbit
contribution, the strength of the Skyrme spin-orbit
parameter $  W  $
has to increase by about 20\% to recover an overall fit to
the single-particle energy data. But the $\ell$-dependence
of the experimental single-particle
energies are better reproduced with the Skyrme spin-orbit shape.
The quality of the Skx and Skxta fit results for single-particle 
energies
around $^{208}$Pb are similar to those we show for $^{132}$Sn.
\begin{table}
\begin{center}
\caption{Contributions to
proton single-particle energies in $^{132}$Sn for
the Skxta and Skx Hartree-Fock calculations. E$^{t}$ is the total
tensor contribution with $\alpha$=93 and $\beta$=94, with the
zero-range tensor interaction contribution for $\alpha_{t}$=60
and $\beta_{t}$=110 shown in brackets.
E$^{so}$ is the spin-orbit potential contribution
to the energy. The total for Skxta
$  E^{t}_{kp}+E^{t}_{kn}+E^{so}  $ is compared with the
spin-orbit for Skx.}
\begin{tabular}{|l|r|r|r|r|r|r|}
\hline
$  n\ell _{j}  $ & $  E^{t}_{kp}  $ &  $  E^{t}_{kn}  $  & $  E^{so}  $ 
 & total
& $  E^{so}  $ \\
 & Skxta &  Skxta  & Skxta  & Skxta
& Skx \\
      & (MeV) & (MeV) & (MeV)  & (MeV) & (MeV) \\
\hline
\hline
 $  0g_{7/2}  $  & -0.723 (-0.476) & -0.828 (-0.984)  &  3.20 &  1.65 & 
 2.64  \\
 $  1d_{5/2}  $  &  0.117 (-0.079) &  0.089 (0.094) & -0.83 & -0.62 & 
-0.68  \\
 $  1d_{3/2}  $  & -0.181 (-0.121) & -0.158 (-0.171)  &  1.34 &  1.00 & 
 1.11  \\
 $  0h_{11/2}  $ &  0.640 (-0.421) &  0.864 (1.034)  & -3.84 & -2.34 & 
-3.19  \\
\hline
\end{tabular}
\end{center}
\end{table}
\begin{figure}
\scalebox{0.5}{\includegraphics{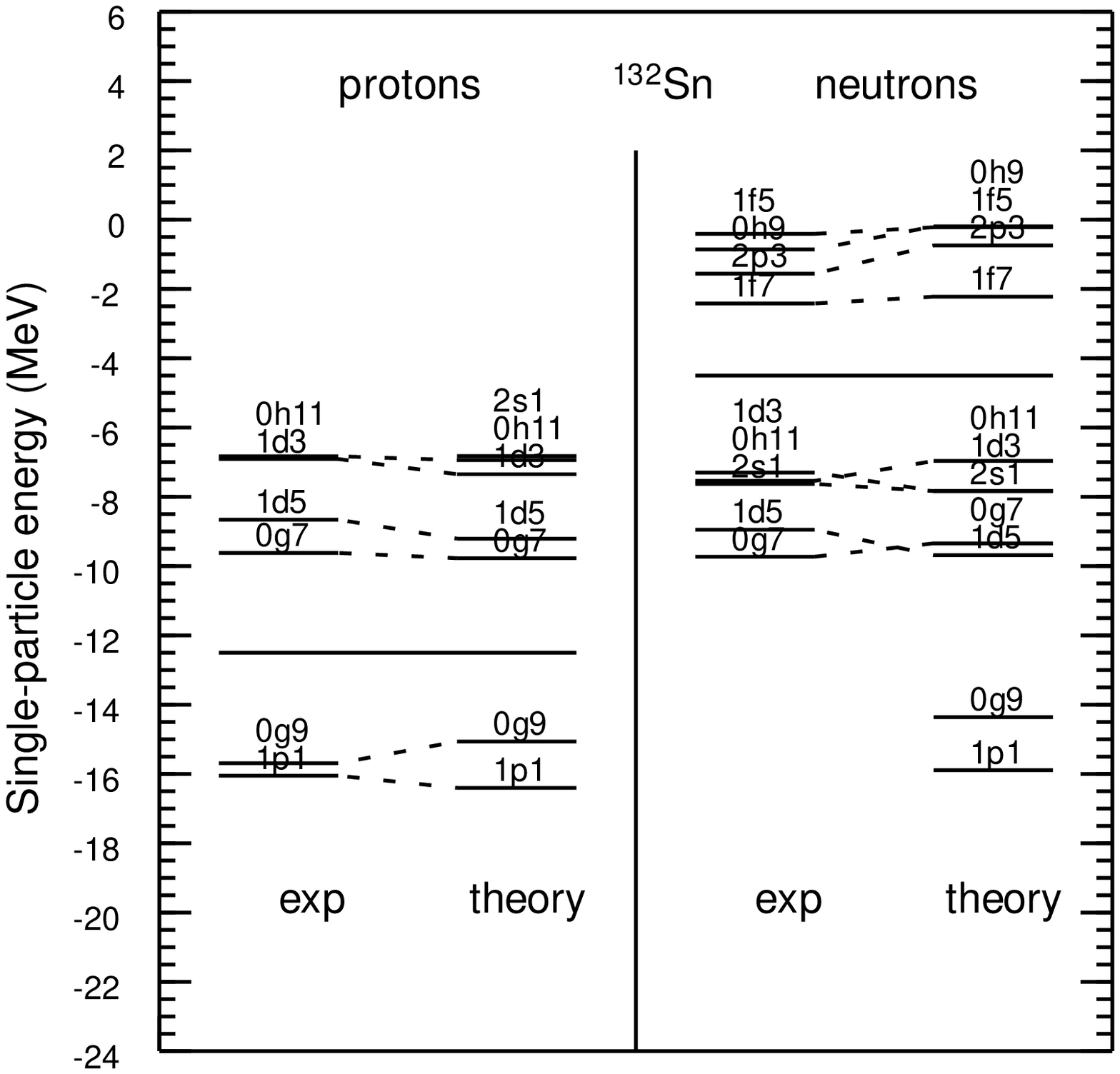}}
\caption{Comparison of experimental and theoretical single-particle
energies in $^{132}$Sn for the Skx interaction.}
\label{(2)}
\end{figure}
\begin{figure}
\scalebox{0.5}{\includegraphics{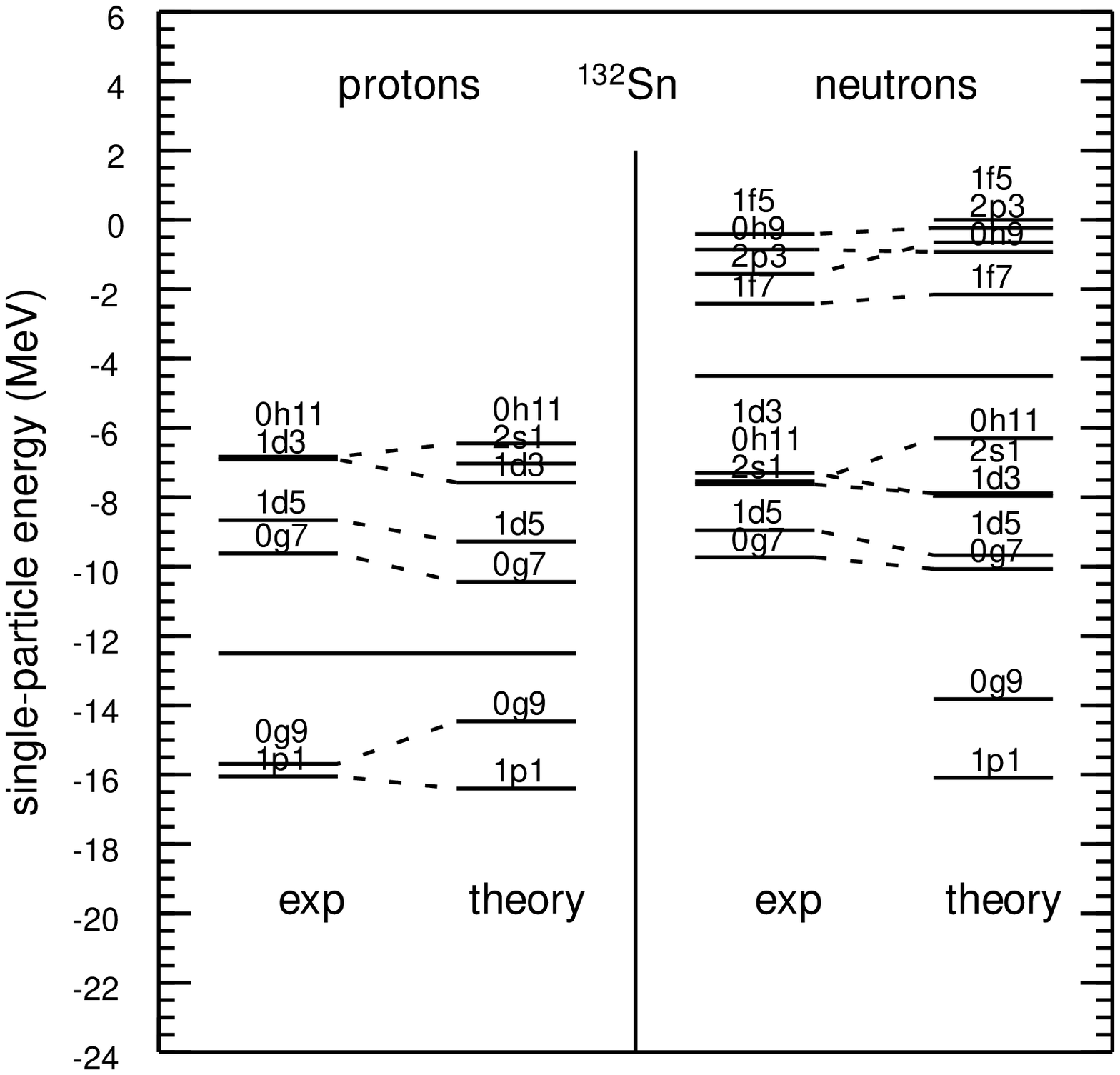}}
\caption{Comparison of experimental and theoretical single-particle
energies in $^{132}$Sn for the Skxta interaction.}
\label{(3)}
\end{figure}
\begin{figure}
\scalebox{0.5}{\includegraphics{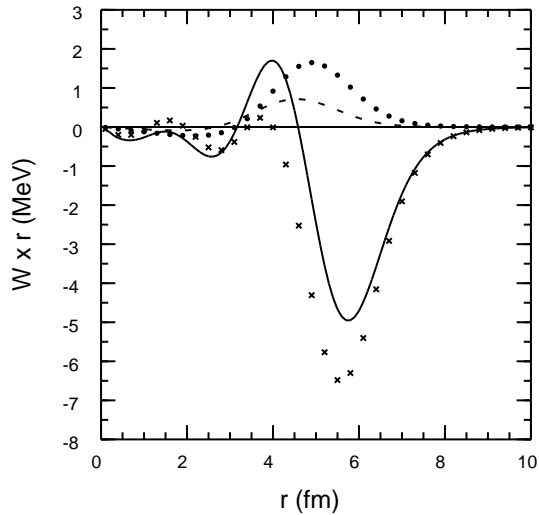}}
\caption{Spin-orbit potentials for protons in $^{132}$Sn for the
Skxta interaction: the Skyrme spin-orbit potential (crosses);
the zero-range
tensor contributions from the core protons (dashed line)
and core neutrons (dotted line); and the total (solid line).
The total also includes the smaller $\alpha_{c}$ and $\beta_{c}$ terms 
from the
exchange part of the central interaction.}
\label{(4)}
\end{figure}

Given that $\beta_{t}$=110 is needed to reproduce the shift of the
$  0g_{7/2}  $-$  0h_{11/2}  $ gap going from $^{114}$Sn to $^{132}$Sn,
we next fix $\beta_{t}$=110 and include $\alpha_{t}$ in the Skyrme fit 
to data.
We recover a good fit of $\chi^{2}$=0.63 with $  \alpha _{t}=-118  $
for a parameter set we call Skxtb.
In general we find a good fit with values of $\beta_{t}$
in the range of 0 to 110
as long as  $  \alpha _{t} \approx -\beta _{t}  $.
This happens because the proton and neutron contributions then
cancel in the $  jj  $ closed shell nuclei $^{132}$Sn and $^{208}$Pb
giving the good reproduction of the single-particle energies
from the Skyrme spin-orbit shape. The Skxtb single-particle
energies are compared with experiment in Table II. Skxtb
gives a best account of the $^{132}$Sn single-particle energies
and the $  0g_{7/2}-0h_{11/2}  $ gap shift. However, the
absolute single-particles energies in $^{114}$Sn still
differ from experiment by as much as one MeV.

In conclusion, we find that the finite-range tensor
interaction is important for the $  0g_{7/2}-0h_{11/2}  $ gap shift.
However, a zero-range implementation of the tensor interaction
in the Skryme interaction is problematic. The radial form
of the tensor contribution to the spin-orbit potential does
not give a good reproduction of the $\ell$-dependence of the
spin-orbit splittings in $^{132}$Sn and $^{208}$Pb. Reproduction
of the observed $  0g_{7/2}-0h_{11/2}  $ gap shift plus
a good fit to absolute single-particle energies in
$^{132}$Sn and $^{208}$Pb requires $  \beta _{t} \approx 110  $ for the
proton-neutron tensor interaction (consistent
with the $  G  $ matrix value), and $  \alpha _{t} \approx -\beta _{t}  
$.
for the $  T=1  $ tensor interaction between like particles,
which
is opposite in sign to the $  G  $ matrix value of
$  \alpha _{t} \approx 60  $. Although our finite-range calculations
indicate that the zero-range approximation may be adequate,
further investigation is required. Also we need to understand
the role of correlations (coupling to vibrations)
and three-body forces on the
effective tensor interactions in nuclei. The central
and part of the Skyrme
functional also need to be
constrained and extended to reproduce
realistic properties of nuclear matter
\cite{lesinski}. These changes may
lead to different values for $\alpha_{c}$ and $\beta_{c}$ than those
obtained with Skx that
also need to be taken when the tensor interaction
is included.

\vspace{ 12pt}
\noindent 
  {\bf Acknowledgments}
Support for this work was provided from US
National Science Foundation grant number PHY-0555366
and PHY-0456903
and by the Japan U.S. Institute for Physics with
Exotic Nuclei (JUSTIPEN). This work
is supported in part by the JSPS
Core-to-Core Program. We thank K. Bennaceur,
M. Bender and T. Unishi for useful discussions.

\end{document}